\documentclass[sigconf,10pt,nonacm]{acmart}
\renewcommand\footnotetextcopyrightpermission[1]{}
\usepackage{hyphenat}
\usepackage{tikz}
\usepackage{amsmath}

\usepackage[ruled,vlined,norelsize,\languagename,linesnumbered]{algorithm2e}
\usepackage{circledsteps}

\usepackage{subfig}
\usepackage{booktabs}
\usepackage{multirow}

\usepackage{xcolor}

\usepackage{xspace}
\usepackage{amsthm}
\usepackage[normalem]{ulem}
\usepackage{enumitem}

\usepackage{flushend}
\usepackage{ifthen}

\renewcommand{\paragraph}[1]{\vspace{0.1em}\noindent\textbf{#1}}

\begin{document}
%-------------------------------------------------------------------------------

\title{Rethinking PM Crash Consistency in the CXL Era}

\author{João Oliveira, João Gonçalves, and Miguel Matos}
\affiliation{%
\institution{IST, Universidade de Lisboa, INESC-ID}
\country{Portugal}
}

\begin{abstract}

Persistent Memory (PM) introduces new opportunities for designing crash-consistent applications without the traditional storage overheads.
However, ensuring crash consistency in PM demands intricate knowledge of CPU, cache, and memory interactions.
Hardware and software mechanisms have been proposed to ease this burden, but neither proved sufficient, prompting a variety of bug detection tools.

With the sunset of Intel Optane comes the rise of Compute Express Link (CXL) for PM.
In this position paper, we discuss the impact of CXL’s disaggregated and heterogeneous nature in the development of crash-consistent PM applications, and outline three research directions: hardware primitives, persistency frameworks, and bug detection tools.

\end{abstract}

\maketitle

\section{Introduction}
\label{sec:intro}

Persistent Memory (PM) is a transformative technology that enables efficient data persistence without the overheads of traditional storage systems.
By providing byte-addressable, cache-coherent access to durable memory, PM has enabled novel application architectures that blur the boundary between memory and storage~\cite{pmemkv, SplitKV, TurboHash, RECIPE, CCEH, FAST-FAIR, PACTree, apex, wipe, HeuristicDB, MadFS, ZoFS, pmemcached, Montage, J-NVM, DINOMO}.
Unfortunately there are no free lunches and fully unlocking the PM potential introduces significant complexity - developers must carefully reason about cache coherence, memory ordering, and crash consistency semantics that differ fundamentally from conventional memory models.

The bulk of these complexities stem from the many optimization layers that are required to make the hardware fast.
On the one hand, since the cache is orders of magnitude faster than PM (and DRAM), regular stores to PM mapped addresses are cached and not guaranteed to be immediately persisted. 
On the other hand, modern CPUs may reorder memory operations, such as stores and flushes that target different cache lines.
To ensure data is persisted in the correct order, developers must use the flush and fence instructions to explicitly move data from the cache to PM.
This programming model is very complex to reason about since it poses a significant shift from the traditional DRAM-based model.
In turn this makes it easier for developers to introduce bugs that impact program correctness and performance.

To mitigate these difficulties the industry and academy have proposed multiple hardware and software mechanisms.

On the hardware front, Intel introduced extended Asynchronous DRAM Refresh (eADR) in 2021 to simplify the persistence model by making caches part of the persistent domain~\cite{eADR,Intel}.
Even though eADR is particularly effective at simplifying the programming model, and preventing PM bugs caused by the reordering and stalling of stores, it does not solve every PM bug.
For example, atomicity bugs cannot be solved via eADR alone, although one could argue that it simplifies the problem.
In these atomic operations, data must be persisted with all\hyp{}or\hyp{}nothing semantics, potentially across cache line boundaries, which eADR does not provide.

On the software side, Intel PMDK~\cite{pmdk} provides high-level abstractions such as crash\hyp{}consistent transactions, and PM object management.
Other frameworks have emerged~\cite{Montage,Mnemosyne,mangosteen,puddles} proposing different abstractions and trade-offs. 
These frameworks cannot cover all possible use cases and, as any software, their implementations can contain errors~\cite{Agamotto, Mumak, PMDebugger, XFDetector} or be misused by developers~\cite{PMDebugger, PMTest}.

Since neither approach proved sufficient, the research community did extensive research on PM bug detection tools as a last line of defense in identifying correctness and performance issues~\cite{Mumak,Yat,Jaaru,Yashme,Witcher,PMTest,PMDebugger,XFDetector,Agamotto,Durinn,PMRace} in PM programs.

Despite this ecosystem, Intel discontinued Optane PM~\cite{pmcanceled}.
Meanwhile, as the Compute Express Link (CXL) standard stabilizes, its PM capabilities promise to be an appealing alternative.
Wheres Optane formed a tightly coupled Intel-only ecosystem with all capabilities in a single unified system, CXL is an open multi-vendor standard with support for disaggregated and heterogeneous systems.
While this heterogeneous and disaggregated nature opens the door for novel designs~\cite{post-optane}, it also greatly expands the fault model.
This means that developers now also need be concerned with partial failures, disaggregation, and hardware heterogeneity.

Perhaps informed by Optane's lessons, CXL attempts to preemptively address some of the common PM programming pitfalls by introducing the Global Persistent Flush (GPF), a mechanism which, similarly to eADR, extends the persistent domain to the cache.
However, the heterogeneous and disaggregated nature of CXL means that GPF's guarantees are much weaker than eADR's, with GPF only ensuring persistence in a limited set of scenarios.

In this position paper we examine how CXL's expanded scope amplifies the existing PM challenges.
In particular, we analyze three key questions facing the systems community:
\begin{itemize}[leftmargin=*]
\item What primitives and abstractions are needed to enable correct and efficient PM programming in disaggregated heterogeneous environments?
\item How should frameworks evolve to handle heterogeneous PM access patterns and failure modes?
\item What novel bug detection approaches are required for distributed PM architectures?
\end{itemize}

\section{Background}
\label{sec:background}

In this section we outline PM Semantics, reason about common PM bugs, and discuss the efforts done at the software and hardware level to prevent or detect them.

\subsection{Persistent Memory Semantics}

\begin{figure}[t]
\includegraphics[width=0.9\columnwidth]{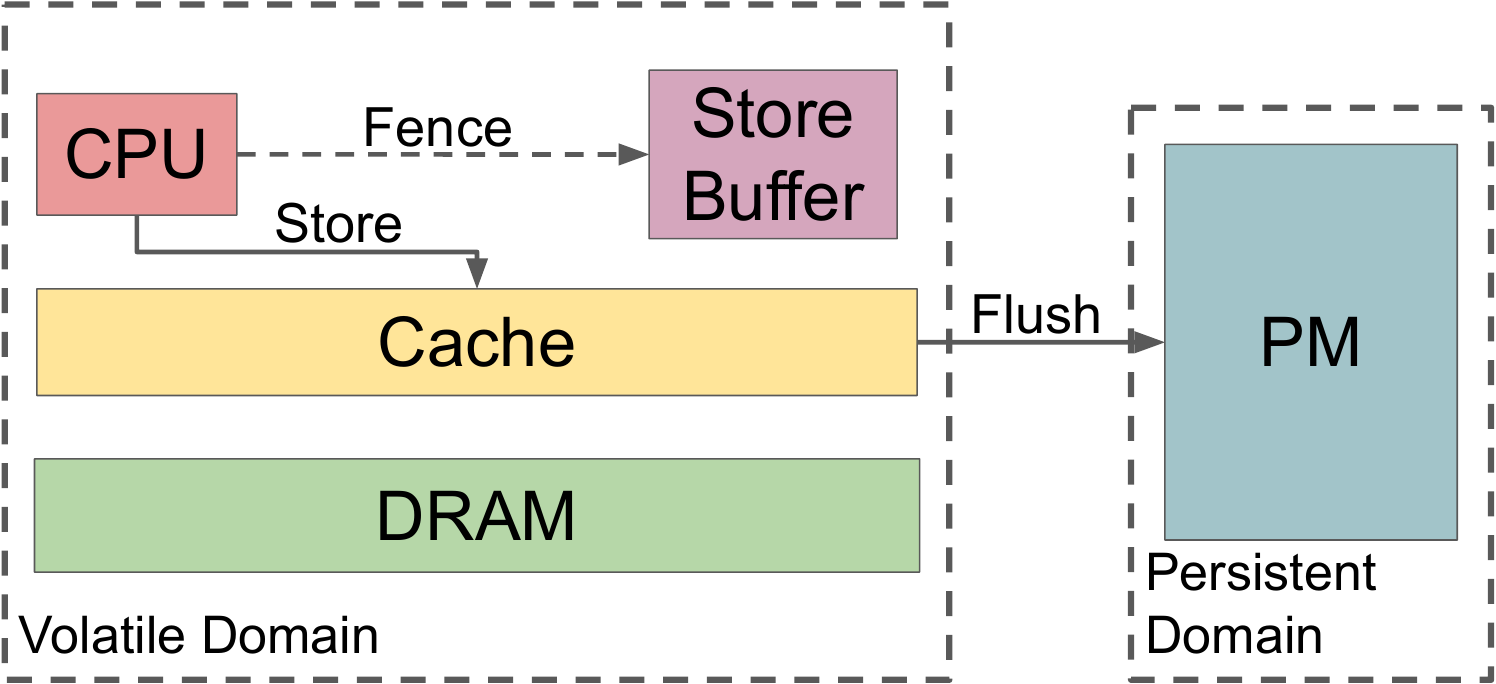}
\caption{A simplified visualization of PM semantics.}
\label{fig:architecture}
\end{figure}

Figure~\ref{fig:architecture} presents an overview of the memory architecture.
When developing PM applications, programmers must reason about the persistency of data.
Stores are considered persistent once they reach a non-volatile domain, which is usually the PM device.
Persistent stores are guaranteed to be visible after a crash (e.g.: a power failure).
Everything else is in the volatile domain, which usually includes the CPU's registers and buffers, the cache, and traditional volatile memory.
Data in the volatile domain is lost after a crash.

PM applications must carefully move data from the volatile domain to the persistent domain. 
This task is error-prone and far from trivial since when working at the code-level there is no clear distinction between volatile and PM accesses -- as in other \emph{mmapped} abstractions, assigning a value to a variable, for instance, \emph{looks} exactly the same regardless of whether that variable's address is in the volatile or PM domain.

On a more conceptual level, there are two key reasons for these difficulties.
First, and for performance reasons, stores are not immediately made durable and remain on the volatile cache.
Although cache-lines may be evicted and written-back to the persistent domain, this happens arbitrarily based on cache-eviction policies, and outside the explicit control of the developers.
Instead, to explicitly move data to the persistent domain, developers need to use specific flush instructions.

Second, modern CPUs can stall and reorder memory operations to amortize their comparatively high costs, via the store buffer.
This reordering includes the aforementioned  flush instructions, which means that they alone are not enough to guarantee explicit persistency.
To address this, the fence family of instructions can be used to guarantee all pending memory operations, specifically, PM stores and their respective flushes, are completed before the execution resumes.
This ensures an order between the cache-lines flushed before and after the fence.
When flush and fence instructions are properly employed, developers can guarantee data is persisted in the correct order.

\subsection{Persistent Memory Bugs}

Given the semantics above, we now discuss simplified versions of commonly encountered PM bug patterns that stem from an incorrect usage of the PM instructions.
Our goal here is not to provide a complete bug taxonomy, but instead provide a few examples, required to support our discussion. 
For a thorough description of PM bug patterns, we refer the reader to Mumak~\cite{Mumak} and Durinn~\cite{Durinn} for a single\hyp{}threaded and concurrency-focused perspective, respectively.

\paragraph{Ordering bugs.}
Algorithm~\ref{alg:simple_bug}, implements a hypothetical put operation in a PM key\hyp{}value store that is organized in buckets (such as P-CLHT~\cite{RECIPE}). 
It iterates through each bucket and, when it finds an empty bucket, sets the new value and then the key.
Our developer is aware of PM ordering issues and hence stores the value before the key in an attempt to ensure that a crash in between both stores will only result in an empty key\hyp{}value pair, which maintains crash-consistency.
However, stores to PM can be reordered and this application still contains a bug since the key might be persisted before the value, followed immediately by a crash.

\begin{algorithm}
\small
\SetNoFillComment
\DontPrintSemicolon
\SetKw{KwIn}{in}

\For{bucket \KwIn buckets} {
\If{bucket = $\bot$} {
value $\leftarrow$ NEW\_VALUE \;
bucket $\leftarrow$ NEW\_KEY \;
} 
}

\caption{Example of an ordering bug}\label{alg:simple_bug}
\end{algorithm}

\paragraph{Atomicity bugs.}
Algorithm~\ref{alg:atomicity_bug}, implements an operation to atomically swap two PM objects.
Our developer carefully flushes and fences both stores.
However,  even though the fence guarantees that, after it executes, all outstanding memory operations are executed, nothing prevents the hardware from (partially) persisting outstanding data before the flush executes.
Thus, a crash after line 1 and before line 4 can result in the loss of one of the objects and an atomicity violation. 

\begin{algorithm}
\small
\SetNoFillComment
\DontPrintSemicolon
\SetKw{KwFlush}{flush}
\SetKw{KwFence}{fence}
objA, objB $\leftarrow$ objB, objA \tcp*{swaps objA and objB}
\KwFlush objA\;
\KwFlush objB\;
\KwFence\;

\caption{Example of an atomicity bug}\label{alg:atomicity_bug}
\end{algorithm}

\paragraph{Concurrency bugs.}
The combination of PM and concurrency introduces a particular pattern of bug, were one thread performs a store and another reads the value of the store before it is persisted.
This combination of data race and PM ordering bug has been studied in the literature~\cite{PMRace,Durinn} and can result in data loss and corruption.

\paragraph{Performance bugs.}
Performance bugs encompass a wide range of issues that negatively impact performance.
These bugs originate from two main causes. 
First, flushing and fencing have a non-negligible cost and can impact access locality thus employing these instructions more often than necessary degrades performance.
Second, despite having comparable performance, PM's access speed is slower than DRAM, specially for writes.
This means that storing volatile data in PM, for instance when some fields of a structure do not need to be persisted, can lead to loss of performance.

\subsection{PM Frameworks and Tools}

As we outlined above, reasoning about PM semantics and programming PM applications is complex.
To make this complexity more manageable, many software frameworks have been proposed to reduce the risk of introducing PM bugs.
Arguably the most proeminent framework is Intel PMDK~\cite{pmdk} which provides abstractions for PM transactions, object lifetime semantics, and memory management among other useful abstractions.
PMDK is closely coupled with ndctl~\cite{ndctl}, a utility for configuring PM regions.
Other frameworks such as Montage~\cite{Montage} proposes an epoch-based PM model. 

These frameworks allow developers to avoid some of the bugs described above, by mapping the application's semantics to the correct abstraction.
For example, to solve the atomicity bug, a developer could leverage the crash\hyp{}consistent transaction abstraction offered by PMDK which, when used correctly, automatically ensures all\hyp{}or\hyp{}nothing semantics.
Note however that this is not a fool\hyp{}proof plan, since even industry grade frameworks contain PM bugs in their design and implementation~\cite{Agamotto,Mumak}.

Based on the recognition that PM programming is hard, even when relying on frameworks, many PM bug detection tools have been proposed~\cite{,Yat,Jaaru,Yashme,Witcher,PMTest,PMDebugger,XFDetector,Agamotto,Mumak,Durinn,PMRace}.
These tools analyze PM applications and search for potential PM bugs, including the ones we discussed earlier.

\subsection{Making the Cache Persistent}

Part of the difficulties of correctly using PM stem from the volatile nature of the cache.
Thus, the idea of extending the persistency domain to include the CPU cache comes rather naturally.
Putting this simple idea into practice is however quite challenging since one cannot simply make all operations in the cache persistent as this would have catastrophic consequences for performance.
The practical alternative is to ensure that dirty cache\hyp{}lines are persisted in case of a crash, removing the need for explicit flush instructions.
This is the approach taken by Intel's extended Asynchronous DRAM Refresh (eADR)~\cite{eADR} which flushes the CPU cache to the PM device upon a crash.

This design has two positive consequences:
i) it might yield better performance since some costly operations are no longer necessary and as a consequence the amount of undesired cache invalidations is also reduced, and
ii) it greatly simplifies the PM programming model, automatically preventing certain PM bugs.

\paragraph{Persistent Cache in Practice.}
Let us now briefly discuss the impact of this mechanism by analyzing its impact on the bug patterns exemplified above.
Recall that the bug in Algorithm~\ref{alg:simple_bug} occurs when the persistency of a store is reordered.
This is because data is only guaranteed to be persistent once it is flushed and, in this case, data may be flushed in an arbitrary order.
By having the cache in the persistent domain, data is made persistent as soon as it reaches the cache, i.e: in program order, thus preventing this bug type.
Furthermore, since dirty cache-lines are flushed to PM upon a crash the application can forgo the use of flush and fence instructions and their associated costs.
This best case scenario yields a simplified programming model and better performance.
However, it does not solve other common issues.

Recall the example in Algorithm~\ref{alg:atomicity_bug} which requires the atomic swap of two PM-backed objects.
Although a persistent cache can ensure that the program order is respected, if the application crashes in the middle of the execution one object may be lost,  which results in the same bug pattern.
In this case one must use a software-based solution such as crash-consistent transactions.

In practice a mechanism such as eADR is costly since it depends on specialized hardware and batteries and therefore developers should not assume that it is universally available.

\section{Compute eXpress Link}
\label{sec:gpf}

CXL is an open standard interconnect, which aims for high\hyp{}performance communication between CPUs, memory, and hardware accelerators.
It is built on top of the PCIe interface, extending the well-known block\hyp{}based protocol (i.e: CXL.io), with two others CXL.cache and CXL.mem.
CXL.io must be implemented by all CXL devices, since it is used for device discovery and configuration.
CXL.cache empowers a CXL device to cache system memory, via the MESI coherence protocol~\cite{MESI}.
CXL.mem allows the host CPU to access a device's memory via a load/store interface.
These protocols can be combined into three different device types:
\textbf{Type 1} implement the CXL.io and CXL.cache protocols, and include, hardware accelerators with coherence requirements, for example smart NICs.
\textbf{Type 2} implement all three protocols, and include hardware accelerators with memory requirements as well as cache\hyp{}coherence like FPGAs.
\textbf{Type 3} are of particular interest to us since they implement the CXL.io and CXL.mem protocols, including PM devices.
Type 3 devices implement several pertinent properties for PM development, namely;

\noindent\textbf{Byte\hyp{}addressable.} CXL.mem allows PM to expose its memory in a byte\hyp{}addressable manner via loads/stores.

\noindent\textbf{Cacheable.} Type 3 device memory can be cached by the host CPU, similarly to data directly accessed via DDR, and in contrast to traditional PCIe block\hyp{}based accesses.

\noindent\textbf{Sharable.} CXL.mem shared memory allows PM to be accessed from multiple CPUs, providing efficient multi\hyp{}host cache\hyp{}coherent shared PM.

\subsection{Global Persistent Flush}

Similarly to Intel's Optane eADR, the CXL standard describes a mechanism to extend the persistent domain to the cache called Global Persistent Flush (GPF).
GPF functions in two distinct sequential phases. 
Phase 1 is responsible for flushing all data from the cache to the persistent device, including the cache of the host CPU, and the caches of other CXL devices.
After the host ensures all caches have been flushed to the persistent device, it can initiate the second phase.
Phase 2 ensures data in volatile buffers within the persistent device is properly flushed to a persistent medium.

Due to the distributed nature, GPF's Phase 1 may fail due to a number of reasons such has having insufficient energy to flush of all caches or due to communication issues with the devices.
This can be detected and logged by the persistent device to be handled upon recovery, although this recovery process is not described in the standard.

\paragraph{Energy.}
A successful execution of the GPF requires, among other things, that system designers take into account specific design goals, calculate the energy requirements for the worst\hyp{}case scenario, and ensure enough power is available to perform the GPF in case of a power-failure.
The standard provides some high-level guidance for this calculation, based on cache\hyp{}size, PM bandwidth and other factors.
Although, at this time it is not clear how much the network topology will impact this calculation, we can surmise that the host and device physical placement in a specific network has the potential to eclipse the energy costs required to successfully complete the GPF.
For example, a host and a PM device, placed several hops apart, in a large scale datacenter, will have much higher power requirements compared to a closely coupled system, where both are placed in the same rack.

\paragraph{Communication.}
Although lack of power is the main concern outlined in the CXL standard, there are other issues that can result in the failure Phase 1.
The most prominent is failures in the communication between the host CPU and the PM device.
Then, irrespectively of the energy reserves, no data can be flushed to the PM device.
Given the complex topologies predicted by the standard, this is likely to become problematic as larger deployments make it into production.

\paragraph{eADR versus GPF.}
When compared to the much simpler, single-host deployment of eADR, which limits energy costs for a complete flushing of dirty cache lines, GPF faces a much harder challenge.
Its implementations must support a heterogeneous system, whose energy requirements may range from negligible to insurmountable. 
Furthermore, the disaggregated nature of CXL means that the communication between CXL hosts and PM devices varies widely, and is more prone to faults.
For example, a CXL switch connecting a host to a PM device is arguably more likely to fail (since it is more complex) than the motherboard bus connecting Intel's Optane PM to the CPU.

\section{Discussion}
\label{sec:discussion}

We now examine the implications of CXL's disaggregated and heterogeneous nature in the development of PM applications.
In regard to the disaggregated nature of CXL, host CPUs and PM devices may be organized in a plethora of topologies containing multiple entities coordinated via a distributed cache\hyp{}coherent mechanism.
One could reason that the sheer breadth of possible topologies for a CXL based system calls for specialized memory primitives better suited for each topology.
The open nature of CXL, coupled with its wide adoption by manufacturers, means that heterogeneity is all but guaranteed in the future of CXL PM devices.
The interfaces by which host CPUs and PM devices communicate (setup, data accesses, and recovery) are well\hyp{}defined under the standard, however, the inner workings of each particular PM device are up to the manufacturer.
This can have impact performance and safety in the application's design and is another axis that developers might need to take into account.

Based on the lessons learned with Optane PM, and consider\-ing CXL's expanded scope, we identify three critical research questions that warrant discussion by the community.

\subsection{What memory primitives are needed?}

As we have outlined in \S\ref{sec:background}, PM support required the introduction of specialized low-level instructions to flush the cache.
These instructions give developers explicit control over persisting data to the PM device, which facilitates the creation of crash\hyp{}consistent PM applications.
With the advent of CXL PM support, could a similar story repeat itself?

Let us analyze another, perhaps less noticed change to Intel's architecture in response to Optane PM.
In the 3rd generation of Intel's Xeon Scalable Processors, \texttt{clwb}, a instruction for writing\hyp{}back a cache\hyp{}line without invalidating it (in certain circumstances) was introduced~\cite{rudoff2017persistent}.
Before this moment, explicitly flushing a cache\hyp{}line, came with the immediate cost of the instruction, as well as the future cost of a cache\hyp{}miss on subsequent accesses to the same cache\hyp{}line. 

According to the current CXL 3.2 standard, CXL.mem devices engaged in cache\hyp{}coherent memory sharing (HDM-DB), cannot explicitly flush a cache\hyp{}line without invalidating it, or forcing a change in the cache\hyp{}line state. 
Both options imply the costs of fetching data from the PM device (which might be remote) during subsequent accesses to the same cache\hyp{}line.
This can lead to undesired performance degradation and variability as applications are deployed in different configurations. 
We believe that the research community should study the feasibility and implications of extending the CXL standard, similarly to what has been done with \texttt{clwb}, to ensure cache\hyp{}lines can be written\hyp{}back without eviction.

Orthogonally to these considerations, the CPU cache eviction policy can still move data from the cache to the persistency domain at any point in time.
Since the persistency domain is now potentially distributed across multiple devices boundaries, this opens another layer of complexity.
On the performance side, programming in a cache conscious manner might become even more important, while on the correctness side one must reason about a distributed persistency domain that now spans multiple failure boundaries.
Is the traditional load/store interface still the right abstraction for expressing distributed persistency semantics?

\subsection{What frameworks are needed?}

One could argue that the availability of frameworks like PMDK~\cite{pmdk} and ndctl~\cite{ndctl} contributed to the adoption of Optane PM.
PMDK offers easy and programmatic configuration of PM pools, object management, crash\hyp{}consistent transactions, while ndctl provides a handy method to configure PM devices.
However, existing frameworks were designed primarily for single-host scenarios with well-defined persistence boundaries and may not cope with CXL's distinguishing characteristics such as resource pooling and multi-host sharing. 
It is easy to see that having frameworks that support higher level abstractions while leveraging CXL PM features is a requirement for the standard to be widely adopted by developers and system designers alike.

Part of this work is already underway, with ndtcl now including a \texttt{cxl} utility for managing CXL memory regions, with special support for PM~\cite{pmdk_cxl}.
Furthermore, the maintainers of PMDK claim that most applications designed for Optane PM can run unmodified on CXL PM~\cite{pmdk_cxl}.

For CXL PM systems that are purely local and hence aligned with the Optane PM model, this transition can indeed be smooth. 
However, the disaggregated nature of CXL, motivates the need for new abstractions in order to take full advantage of the technology.

For example, multi\hyp{}host shared\hyp{}memory enables the development of multi\hyp{}host PM applications, which in turn require appropriate abstractions.
As far as we can tell, PMDK's abstractions, like transactions, are not designed for such environments.
The future of CXL PM development requires the creation of apt transactions, that can enable all facets of the standard, and as such, we believe that more research is required in this direction. 

\subsection{What bug detection tools are needed?}

Similarly to what happened with Optane PM, regardless of the quality and expressiveness of the low-level primitives and high-level abstractions that the community ends up building, we still need tools to assist developers in finding PM-specific bugs.
The PM bug detection tools of the past are not capable of reasoning about the inherent properties of disaggregated systems, including CXL PM.
CXL PM expands the fault model, with potential to introduce novel PM bug patterns, which require new techniques to detect them.

While GPF might rule out some bugs patterns in specific scenarios, as we discused in \S\ref{sec:gpf}, it does not solve all problems since it is prone to failure due to energy requirement or communication constraints.
Furthermore, even if a system is designed and deployed in such a way that GPF is always able to complete, the mechanism can not prevent every PM bug, such as the atomicity bug pattern outlined in Algorithm~\ref{alg:atomicity_bug}, which requires a software solution.
Additionally, even if PM frameworks might offer abstractions that prevent some classes of bugs, the frameworks can have bugs themselves, or be employed incorrectly as we have witnessed in the past.

For these reasons we believe that GPF must not be taken as a catch\hyp{}all method to get rid of crash-consistency related problems, and should instead be seen as a layer of protection in the fight against catastrophic post\hyp{}crash behaviors, such as data loss or corruption.

We argue therefore that the systems community needs to develop new bug detection tools aware of the specificities of CXL such as potentially inconsistent ordering guarantees across a distributed persistence domain.

\section{Conclusion}
\label{sec:conclusion}

With this position paper we aim to spark discussion on developing the right combination of hardware mechanisms, software frameworks, and development tools to make CXL PM's benefits accessible to a wide range of developers while containing its complexity.
The lessons learned from Optane provide valuable insights, but CXL's broader scope demands fresh approaches to these recurring challenges.

%-------------------------------------------------------------------------------

\bibliographystyle{ACM-Reference-Format}
\bibliography{bib_DB}

\begin{acks}
This work was supported by Fundação para a Ciência e a Tecnologia (FCT) under PhD scholarships 2021.07401.BD and 2024.01891.BD, and research project grant PTDC/CCI-COM/4485/2021 (Ainur). 
\end{acks}
\balance

%%%%%%%%%%%%%%%%%%%%%%%%%%%%%%%%%%%%%%%%%%%%%%%%%%%%%%%%%%%%%%%%%%%%%%%%%%%%%%%%
\end{document}